\documentstyle [12pt]{article}                                                
\begin{document}                                                              
\title{ Strings and branes with a modified measure                           
\protect\\  } \author{E.I. Guendelman                                         
\\{\it Physics Department, Ben-Gurion University, Beer-Sheva                  
84105, Israel}}                                                               
                                                                              
\maketitle                                                                    
\bigskip                                                                      
\begin{abstract}                                                              
In string theory, the consequences of replacing the measure of 
integration $\sqrt{-\gamma}d^2 x$ in the Polyakov's action by $\Phi d^2 x$
where $\Phi$ is a  density built out of degrees of freedom 
independent of the metric $\gamma_{ab}$ defined in the string are 
studied. The string tension appears as an integration constant of the 
equations of motion. The string tension can change in different parts 
of the string due to the coupling of gauge fields and point particles 
living in the string. The generalization to higher dimensional extended
objects is also studied. In this case there is no need of a fine tuned 
cosmological term, in sharp contrast to the standard formulation of the
generalized Polyakov action for higher dimensional branes.      
                                   
\end{abstract}

\section{Introduction}

String and brane theory \cite{Yuval} have appeared as candidates for unifying all interactions 
of nature. One aspect of string and brane theories seems to many not quite appealing
however: this is the introduction from the begining of a fundamental scale,
the string or brane tension.
The idea that the fundamental theory of nature, whatever that may be, 
should not contain any fundamental scale has attracted a lot of attention.
According to this point of view, whatever scale appears in nature, must not
appear in the fundamental lagrangian of physics. Rather, the appearence 
of these scales must be spontaneous, for example due to boundary conditions
in a classical context or a process of dimensional transmutation to give an
example of such effect in the context of quantum field theory.

Also, in the context of gravitational theory, the idea that Newton's
constant may originate from a phenomenon of spontaneous symmetry breaking has
inspired Zee \cite{Zee} and others to build models along these lines. Furthermore, 
it has been shown that Zee's induced gravity can in turn be obtained from a
theory without any fundamental scale and manifest global scale invariance
\cite{ind-mea}. In such an approach, when integrating the equations of motion, we introduce 
an integration constant which is responsible for the ssb of scale invariance,
as described in the model studied in Refs. 4,5 6, which was shown to be 
connected, after the ssb, to the Zee model in Ref.3.

The model of Refs.3,4,5,6  is based on the possibility of replacing 
the measure of integration $\sqrt{-\gamma}d^{D}x$, by another one,               
$\Phi d^{D}x$, where $\Phi$ is a density built out of degrees of freedom      
independent of the metric $\gamma_{ab}$ . Such possibility was studied 
in a general context \cite{modmes}, not related to scale invariance also.

Here we want to see what are the consequences of doing something similar in 
the context of string theory. As we will see, string theories or more 
generally brane theories without a fundamental scale are possible if the 
extended objects do not have boundaries (i.e., they are closed).

In the context of the formalism for extended object proposed here, 
if there are boundaries, we  require
the coupling of a gauge field that lives in the brane                                         
to a lower dimensional object (a point particle in the case of a string) that
defines the boundary, since then the equations of motion allow us to end the 
extended object at this boundary. The coupling constant of the gauge field 
to the lower dimensional object defines in this case a fundamental scale of 
the theory. The scales of the theory appear therefore as integration 
constants in the case of closed extended objects or through the physics of 
the boundaries of these extended objects.

\section{String theories with a modified measure}

The Polyakov action for the bosonic string is \cite{Poly}

 \begin{equation}
 S_{P}[X,\gamma_{ab}] =   -T \int d\tau d\sigma \sqrt{-\gamma}
\gamma^{ab} \partial_{a} X^{\mu}\partial_{b} X^{\nu} g_{\mu \nu}                              
 \end{equation}
                                                
Here $\gamma_{ab}$ is the metric defined in the $1+1$ world sheet 
of the string and $ \gamma = det(\gamma_{ab})$. $g_{\mu \nu}$ is the 
metric of the embedding space. $T$ is here the string tension, a 
dimensionfull quantity introduced into the theory, which defines a
scale .

We recognize the measure of integration
$d\tau d\sigma \sqrt{-\gamma}$ and as we anticipated before, we want 
to replace this measure of integration by another one which does 
not depend on $\gamma_{ab}$ .  

If we introduce two scalars (both from the point of view of the $1+1$
world sheet of the string and from the embedding $D$-dimensional universe)
$\varphi_{i}$, $i=1,2$, we can contruct the world sheet density

\begin{equation}                                                              
\Phi =  \varepsilon^{ab}  \varepsilon_{ij}                   
\partial_{a} \varphi_{i} \partial_{b} \varphi_{j}                                      
\end{equation}                                                                
                                                                              
where $\varepsilon^{ab}$ is given by $\varepsilon^{01} = 
-\varepsilon^{10} =1$, $\varepsilon^{00} =                      
\varepsilon^{11} = 0 $ and $\varepsilon_{ij} $ is defined by
$\varepsilon_{12} = -\varepsilon_{21} = 1$, $\varepsilon_{11} =
\varepsilon_{22} =0 $. 

It is interesting to notice that $d\tau d\sigma \Phi = 
2 d \varphi_{1} d \varphi_{2}$, that is the measure of integration
$d\tau d\sigma \Phi $ corresponds to integrating in the space 
of the scalar fields $\varphi_{1}, \varphi_{2}$.

We proceed now with the construction of an action that uses 
$d\tau d\sigma \Phi$ instead of $d\tau d\sigma \sqrt{-\gamma}$. 
When considering the types of actions we can have under these
circumtances, the first one that comes to mind ( a straightforward
generalization of the Polyakov action) is

\begin{equation}
S_{1} = - \int d\tau d\sigma \Phi
\gamma^{ab} \partial_{a} X^{\mu}\partial_{b} X^{\nu} g_{\mu \nu} 
\end{equation} 
          
Notice that multiplying $S_{1}$ by a constant, before boundary or initial
conditions are specified is a meaningless operation, since such a constant 
can be absorbed in a redefinition of the measure fields $\varphi_{1}, 
\varphi_{2}$ that appear in $\Phi$.

The form (3) is however not a satisfactory action, because the variation of 
$S_{1}$ with respect to $\gamma^{ab}$ leads to the rather strong condition

\begin{equation}
\Phi \partial_{a} X^{\mu}\partial_{b} X^{\nu} g_{\mu \nu} = 0
\end{equation}            

If $\Phi \neq 0 $, it means that 
$ \partial_{a} X^{\mu}\partial_{b} X^{\nu} g_{\mu \nu} = 0$, which means
that the metric induced on the string vanishes, clearly not an acceptable
dynamics. Alternatively, if $\Phi=0$, no further information is available, 
also a not desirable situation.
                                                                                                                                                                
To make further progress, it is important to notice that terms that when 
considered as contributions to $L$ in  
                               
\begin{equation}                                                              
S = \int d\tau d\sigma \sqrt{-\gamma}  L                   
\end{equation}                                                                
                                        
which do not contribute to the equations of motion, i.e., such that 
$\sqrt{-\gamma}  L$ is a total derivative, may contribute when we 
consider the same $L$, but in a contribution to the action of the form

\begin{equation}        
S = \int d\tau d\sigma \Phi L
\end{equation} 

This is so because if $\sqrt{-\gamma}  L$ is a total divergence, 
$ \Phi L$ in general is not.

This fact is indeed crucial and if we consider an abelian gauge field 
$A_{a}$ defined in the world sheet of the string, in addition to the 
measure fields $\varphi_{1},         
\varphi_{2}$ that appear in $\Phi$, the metric $\gamma^{ab}$ and the
string coordinates $ X^{\mu}$, we can then construct the non trivial
contribution to the action of the form

\begin{equation}
 S_{gauge} = \int d\tau d\sigma \Phi 
\frac {\varepsilon^{ab}}{\sqrt{-\gamma}} F_{ab}
\end{equation}                                       
 where 
\begin{equation} 
F_{ab} = \partial_{a} A_{b} - \partial_{b} A_{a}
\end{equation}

So that the total action to be considered is now
\begin{equation}   
S = S_{1} +  S_{gauge} 
\end{equation} 

 with $S_{1}$ defined as in eq. 3 and $S_{gauge}$ defined by eqs.7 and 8.

The action (9) is invariant under a set of diffeomorphisms in the space
of the measure fields combined with a conformal transformation of the metric
$\gamma_{ab}$,
\begin{equation} 
\varphi_{i} \rightarrow \varphi_{i}^{'} = \varphi_{i}^{'} (\varphi_{j})
\end{equation}

So that,

\begin{equation}
\Phi \rightarrow \Phi^{'} = J \Phi 
\end{equation} 
where J is the jacobian of the transformation (10)
and 
\begin{equation}
\gamma_{ab} \rightarrow \gamma^{'}_{ab} = J \gamma_{ab}
\end{equation} 
        
The combination $\frac {\varepsilon^{ab}}{\sqrt{-\gamma}} F_{ab} $ is a 
genuine scalar. In two dimensions is proportional to $\sqrt{ F_{ab} F^{ab}}$.

Working with (9), we get the following equations of motion:
From the variation of the action with respect to $\varphi_{j}$

\begin{equation}
\varepsilon^{ab} \partial_{b} \varphi_{j} \partial_{a} (
-\gamma^{cd} \partial_{c} X^{\mu}\partial_{d} X^{\nu} g_{\mu \nu} +
\frac {\varepsilon^{cd}}{\sqrt{-\gamma}} F_{cd} ) = 0
\end{equation}

If $det (\varepsilon^{ab} \partial_{b} \varphi_{j}) \neq 0$, which
means $\Phi \neq 0$, then we must have that all the derivatives of 
the quantity inside the parenthesis in eq.13 must vanish, that is, 
such a quantity must equal a constant which will be determined later,
but which we will call $M$ in the mean time,   

\begin{equation} 
-\gamma^{cd} \partial_{c} X^{\mu}\partial_{d} X^{\nu} g_{\mu \nu} +
\frac {\varepsilon^{cd}}{\sqrt{-\gamma}} F_{cd} = M
\end{equation}  

The equation of motion of the gauge field $A_{a}$, tells us about 
how the string tension appears as an integration constant. 
Indeed this equation is

\begin{equation} 
\varepsilon^{ab} \partial_{b} (\frac {\Phi}{\sqrt{-\gamma}}) = 0
\end{equation}

which can be integrated to give

\begin{equation}
\Phi = c \sqrt{-\gamma}
\end{equation}

Notice that (16) is perfectly consistent with the conformal symmetry
(10), (11) and (12). Equation 14 on the other hand is consistent with 
such a symmetry only if $M = 0$. Indeed, we will check that the equations 
of motion indeed  imply that $M = 0$. In the case of higher dimensional 
branes, the equations of motion require also a very specific value of  $M$,
but in that case, it will be a non vanishing value.

By calculating the Hamiltonian, after dropping boundary terms (this is
totally justified in the case of closed strings) and (only at the end of
the process) using eq.16,  we find that $c$ equals the string 
tension. 

Furthermore, if we couple the gauge field $A_{a}$ to point particles living
in the string, we find that the right hand side of eq.15 is not zero anymore,
but rather a delta function with non vanishing support at the location of the
particle. The solution of the equation will be $\Phi = c_{1} \sqrt{-\gamma} $
to the right of the point particle and $\Phi = c_{2} \sqrt{-\gamma} $ to the
left of the point particle. $ c_{2} -  c_{1} $ will be then the charge of the 
point particle. We obtain the a picture of a string where the tension has 
changed from one region to the other according to the charge that we have 
inserted.

There is the possibility that either $c_{1}$ or $c_{2}$ equal zero. In that
case the string itself starts at the location of the elementary charge. This
picture could be of use in for example a string confinement model of charged
particles. In this case a fundamental scale is introduced, not through the
straightforward introduction of a string tension, but by the introduction of 
the elementary charge of a point particle living in the string, i.e. through
the boundary physics of the string. 

It is very important to notice that in this formulation the string can finish
at a certain definite boundary, in a way that is dictated by the equations
of motion, due to the introduction of point like charges at the boundaries 
of the string. This allows the measure to just vanish when we go beyond the
point like charge.

Now let us turn our attention to the equation of motion derived from the
variation of (9) with respect to $\gamma^{ab}$. We get then,

\begin{equation}
 - \Phi (\partial_{a} X^{\mu}\partial_{b} X^{\nu} g_{\mu \nu}
- \frac {1}{2} \gamma_{ab} \frac {\varepsilon^{cd}}{\sqrt{-\gamma}} F_{cd}) = 0  
\end{equation}  

From the constraint (14), we can solve 
$\frac {\varepsilon^{cd}}{\sqrt{-\gamma}} F_{cd} $ and insert back into 
(17), obtaining then (if $\Phi \neq 0$)

\begin{equation}
\partial_{a} X^{\mu}\partial_{b} X^{\nu} g_{\mu \nu} -
 \frac {1}{2} \gamma_{ab} 
\gamma^{cd} \partial_{c} X^{\mu}\partial_{d} X^{\nu} g_{\mu \nu}
-\frac {1}{2} \gamma_{ab} M = 0
\end{equation}

Multiplying the above equation by  $\gamma^{ab}$ and summing over $a, b$,
we get that $M = 0$, that is the equations are exactly those of the Polyakov action. 
After eq.16 is used, the eq. obtained from the variation of $X^{\mu}$ is
seen to be exactly the same as the obtained from the Polyakov action as well.

\section{Higher Dimensional Extended Objects}  

Let us now consider a $d+1$ extended object, described (generalizing the 
action (9)),

\begin{equation} 
S = S_{d} + S_{d-gauge}
\end{equation}  

where 
\begin{equation}
 S_{d} = -\int d^{d+1} x \Phi                                             
\gamma^{ab} \partial_{a} X^{\mu}\partial_{b} X^{\nu} g_{\mu \nu}
\end{equation}
and 

\begin{equation}
S_{d-gauge} =  \int d^{d+1} x \Phi
\frac {\varepsilon^{a_{1}a_{2}...a_{d+1}}}{\sqrt{-\gamma}} 
\partial_{[a_{1}}A_{a_{2}...a_{d+1}]}
\end{equation} 

and
\begin{equation}
\Phi = \varepsilon^{a_{1}a_{2}...a_{d+1}}\varepsilon_{j_{1}j_{2}...j_{d+1}}
\partial_{a_{1}} \varphi_{j_{1}}....\partial_{a_{d+1}} \varphi_{j_{d+1}}
\end{equation} 

This model does not have a symmetry which involves an arbitrary 
diffeomorphism in the space of the measure fields coupled with a conformal
transformation of the metric, except if $ d=1 $ (eqs. (10), (11), (12)). 
For $d \neq 1$, there is 
still a global
scaling symmetry where the metric transforms as ($\theta$ being a constant),
                  
\begin{equation}                                                              
\gamma_{ab}  \rightarrow   e^{\theta} \gamma_{ab}                              
\end{equation}                                                                

the $\varphi_{j}$ are transformed according to                                 

\begin{equation}
\varphi_{j}   \rightarrow   \lambda_{j} \varphi_{j}                         
\end{equation} 

(no sum on $j$) which means                                                     
$\Phi \rightarrow \biggl(\prod_{j} {\lambda}_{j}\biggr) \Phi \equiv \lambda   
\Phi $

Finally, we must demand that $\lambda = e^{\theta} $  and that the transformation
of $A_{a_{2}...a_{d+1}}$ be defined as

\begin{equation} 
A_{a_{2}...a_{d+1}} \rightarrow \lambda ^{\frac {d-1}{2}}A_{a_{2}...a_{d+1}}
\end{equation}  

Then we have a symmetry. Also no scale is introduced into the theory from 
the beginning. This is apparent from the fact that any constants multiplying
the separate contributions to the action (20) or (21) is meaningless if no 
boundary or initial conditions are specified, because then such factors can 
be absorbed  by a redefinition of the measure fields and of the gauge fields.
Notice that the existence of a symmetry alone is not enough to guarantee that
no fundamental scale appears in the action. For example string theory, 
as usually formulated has conformal symmetry, but the string tension is 
still a fundamental scale in the theory.

Another interesting symmetry of the action (up to the integral of a total 
divergence) consists of the infinite dimensional set of transformations                                        
$\varphi_{j} \rightarrow \varphi_{j} + f_{j} (L)$, 
where $f_{j} (L)$ are arbitrary functions of
 
\begin{equation}
L = - \gamma^{cd} \partial_{c} X^{\mu}\partial_{d} X^{\nu} g_{\mu \nu} +              
 \frac {\varepsilon^{a_{1}a_{2}...a_{d+1}}}{\sqrt{-\gamma}}                   
\partial_{[a_{1}}A_{a_{2}...a_{d+1}]}                     
\end{equation} 

This symmetry does depend on the explicit form of the lagrangian density
$L$ , but only the fact that  $L$ is $\varphi_{a}$ independent.

Now we go through the same steps we went through in the case of the string.
The variation with respect to the measure field $\varphi_{j}$ gives 

\begin{equation}
K^{a}_{j} \partial_{a} (                      
-\gamma^{cd} \partial_{c} X^{\mu}\partial_{d} X^{\nu} g_{\mu \nu}
+ \frac {\varepsilon^{a_{1}a_{2}...a_{d+1}}}{\sqrt{-\gamma}}                    
\partial_{[a_{1}}A_{a_{2}...a_{d+1}]}) = 0
\end{equation}

where 
\begin{equation}                                                              
K^{a}_{j} = \varepsilon^{a a_{2}...a_{d+1}}\varepsilon_{j j_{2}...j_{d+1}}   
\partial_{a_{2}} \varphi_{j_{2}}....\partial_{a_{d+1}} \varphi_{j_{d+1}}      
\end{equation}                 

Since $det(K^{a}_{j})= \frac {(d+1)^{-(d+1)}}{(d+1)!}\Phi^{d}$, it therefore 
follows that for $\Phi \neq 0$, $det(K^{a}_{j})\neq 0$ and

\begin{equation}  
-\gamma^{cd} \partial_{c} X^{\mu}\partial_{d} X^{\nu} g_{\mu \nu} 
+ \frac {\varepsilon^{a_{1}a_{2}...a_{d+1}}}{\sqrt{-\gamma}}
\partial_{[a_{1}}A_{a_{2}...a_{d+1}]} = M 
\end{equation}

where $M$ is some constant of integration. If $d \neq 1$ then $M \neq 0$ as we
will see. Furthermore, under a scale transformation (23), (24), (25),
$M$ does change from one constant value to another.

The variation with respect to the gauge field $A_{a_{2}...a_{d+1}}$ leads to
the equation 

\begin{equation}
\varepsilon^{a_{1}a_{2}...a_{d+1}}\partial_{a_{1}}
\frac {\Phi}{\sqrt{-\gamma}} = 0 
\end{equation}  

which means

\begin{equation}                                                              
\Phi = c \sqrt{-\gamma}                                                       
\end{equation}                                                                

once again. As in the case of the string the brane tension has been generated
spontaneously instead of appearing as a parameter of the fundamental 
lagrangian. Again a simple calculation of the Hamiltonian and using after this
the above equation, we obtain that $c$ equals the brane tension.

The variation of the action with respect to $\gamma^{ab}$ leads to

\begin{equation}
- \Phi (\partial_{a} X^{\mu}\partial_{b} X^{\nu} g_{\mu \nu}               
- \frac {1}{2} \gamma_{ab} 
\frac {\varepsilon^{a_{1}a_{2}...a_{d+1}}}{\sqrt{-\gamma}} 
\partial_{[a_{1}}A_{a_{2}...a_{d+1}]}) = 0 
\end{equation}  

We can now solve for 
$\frac {\varepsilon^{a_{1}a_{2}...a_{d+1}}}{\sqrt{-\gamma}}
\partial_{[a_{1}}A_{a_{2}...a_{d+1}]}$ from equation (29) and then reinsert 
in the above equation, obtaining then, 

\begin{equation} 
\partial_{a} X^{\mu}\partial_{b} X^{\nu} g_{\mu \nu} =
\frac {1}{2} \gamma_{ab} 
(\gamma^{cd}\partial_{c} X^{\mu}\partial_{d} X^{\nu} g_{\mu \nu} + M )
\end{equation} 

This is the same equation that we would have obtained from a Polyakov type
action augmented by a cosmological term.

As in the case of the string, $M$ can be found by contracting both sides of
the equation. For $d \neq 1 $, $M$ is non zero and equal to
\begin{equation}
M = 
\frac{\gamma^{cd}\partial_{c} X^{\mu}\partial_{d} X^{\nu} g_{\mu \nu} (1-d)}{1+d}
\end{equation}

We can also solve for 
$\gamma^{cd}\partial_{c} X^{\mu}\partial_{d} X^{\nu} g_{\mu \nu}$ in terms of 
$M$ from (34) and insert in the right hand side of (33), obtaining,

\begin{equation}
\gamma_{ab} = \frac {1-d}{M}
\partial_{a} X^{\mu}\partial_{b} X^{\nu} g_{\mu \nu}
\end{equation}

Which means that $\gamma_{ab}$ is up to the constant 
factor $\frac {1-d}{M}$ equal
to the induced metric. Since there is the scale invariance (23), (24), (25),
an overall constant factor in the evolution of $\gamma_{ab}$ cannot be 
determined. The same scale invariance means however that there is a field
redefinition which does not affect any parameter of the lagrangian and
which allows us to set $\gamma_{ab}$ equal to the induced metric (at least
if we start from any negative value of $M$), that is,

\begin{equation}
\gamma_{ab} = \partial_{a} X^{\mu}\partial_{b} X^{\nu} g_{\mu \nu} 
\end{equation}

In such case $M$ is consistently given (inserting (36) into (35) or (34)),

\begin{equation} 
M = 1-d
\end{equation} 

Notice that in contrast with the standard approach for Polyakov type actions 
in the case of higher dimensional branes \cite{lambda}, here we do not have to 
fine tune a 
parameter of the lagrangian the brane "cosmological constant", so as to 
force that (36) be satisfied. Rather, it is an integration constant,
that appears from an action without an original cosmological term, which can be
set to the value given by eq. (37) by means of a scale transformation. Such 
choice ensures then that (37) is satisfied (and therefore (36)).
Furthermore, it appears that this 
treatment is more appealing if one thinks of all branes on similar footing, 
since in the approach of this paper they can all be described by a similar 
looking lagrangian, unlike in the usual aproach which discriminates in a 
radical way between strings, these having no cosmological constant 
associated to them, and the higher dimensional branes, which require a
fine tuned cosmological constant.

As in the case of the string, the constant $c$ provides a spontaneously
generated brane tension.
In a way similar to that of the string, we can generate a discontinuity
in such brane tension by coupling minimally the gauge field 
$A_{a_{1}....a_{d}}$ defined in the brane to a current defined in the 
boundary of such a brane, which is a lower dimensional brane.

As in the case of the string, the brane can finish at a certain definite 
boundary, in a way that is dictated by the equations of motion, due to 
the introduction  charges at the boundaries, which are lower dimensional
branes. This allows the measure to just vanish when we go beyond the 
boundaries defined by the lower dimensional brane.

\section{Discussion and Conclusions}
A different approach to the theory of extended objects has been developed
by allowing the integration measure in the action to be independent of the
metric.

In this approach, in the case of closed objects, no scales appear in the
fundamental lagrangian, the brane tension appears as an integration constant.
If coupling to lower dimensional branes are allowed, this coupling introduces a
a fundamental scale. That it, scales are introduced only as the result of 
initial conditions or as the result of the physics of the boundaries of the 
extended object.

Further generalizations and extensions to incorporate supersymmetry should
be studied in order to build a relistic model. The fact that
both strings and branes can be studied with a fundamental action which 
does not contain an explicit cosmological term, in contrast with the usual
treatment, which requires a different cosmological term for every type of 
brane, should be of use when trying to achieve a unified treatment of all
these branes.

One should notice that other authors have also constructed 
actions for branes that do not contain a brane-cosmological term \cite{Dolan}.
Such formulations depend, unlike what has been developed here, on the 
dimensionality, in particular whether this is even or odd, so that it is 
clear that those formulations do not have much relation with what has 
been done here. Yet other approaches \cite{Barc}
to an action without a brane cosmological involve lagrangians with non linear
dependence on the invariant 
$\gamma^{cd}\partial_{c} X^{\alpha}\partial_{d} X^{\beta} g_{\alpha \beta}$,
also a rather different path to the one followed here. For an interesting
analysis of  different possible Lagrangians for extendons 
see \cite{Eiz}.

An approach that has some common features to the one developed here is
that of Ref. \cite{Berg}, where also the tension of the brane is found 
as an integration constant. Here also gauge fields are introduced, but they
appear in a quadratic form rather than in a linear form. Also the scale
invariance discussed there is a target space scale invariance since no 
metric defined in the brane is studied there, i.e. no connection to a 
Polyakov type action, which is known to be more useful in the
quantum theory, is made.

Finally, it will be of use to develop theories along the lines developed here 
not only as candidate unified models for all fundamental interactions, but
also as useful phenomenological tools for the study of confinement of 
quarks.

\section{Acknowledgements} I want to thank C.Castro in particular for
pointing to me the related work in ref. \cite{Berg}. I also want to
thank A.Davidson, A.Kaganovich, J.Portnoy and 
L.C.R. Wijewardhana for discussions.

\end{document}